\documentstyle[12pt,fleqn]{article}
\input epsf
\def\be{\begin{equation}}
\def\ee{\end{equation}}
\def\ba{\begin{eqnarray}}
\def\ea{\end{eqnarray}}
\def\fun#1#2{\lower3.6pt\vbox{\baselineskip0pt\lineskip.9pt
        \ialign{$\mathsurround=0pt#1\hfill##\hfil$\crcr#2\crcr\sim\crcr}}}
\def\la{\mathrel{\mathpalette\fun <}}

\def\eqr#1{{Eq.~(\ref{#1})}}

\begin{document}

\begin{titlepage}
\vspace*{-64pt}
\begin{flushright} {\footnotesize
CERN-TH/2003-105\\
hep-ph/0305074 \\ }
\end{flushright}

\vskip 1.5cm

\begin{center}
{\Large\bf MINIMAL THEORETICAL UNCERTAINTIES \\
\vskip 0.5cm 
 IN INFLATIONARY PREDICTIONS}

\vskip 1cm
{\bf
Daniel J. H. Chung,$^{a,b}$\footnote{E-mail: {\tt Daniel.Chung@cern.ch}}
 Alessio Notari,$^{c}$\footnote{E-mail: {\tt a.notari@sns.it }}
 Antonio Riotto$^{d}$\footnote{E-mail: {\tt antonio.riotto@pd.infn.it}}
}
\vskip .75cm
{\it
$^a$Theory Division, CERN, CH-1211 Geneva 23, Switzerland\\
$^b$ Department of Physics, University of Wisconsin, Madison, WI 53706, USA\\
\vspace{12pt}

$^c$Scuola Normale Superiore, Piazza dei Cavalieri 7, Pisa I-56126, Italy\\
\vspace{12pt}
$^d$INFN, sezione di Padova, via Marzolo 8, Padova I-35131, Italy
\vspace{12pt}
}
\end{center}
\vskip .5cm

\begin{quote}
\noindent
During inflation, primordial energy density fluctuations are created
from approximate de Sitter vacuum quantum fluctuations redshifted out
of the horizon after which they are frozen as perturbations in the
background curvature. In this paper we demonstrate that there exists
an intrinsic theoretical uncertainty in the inflationary predictions
for the curvature perturbations due to the failure of the well known
prescriptions to specify the vacuum uniquely.  Specifically, we show
that the two often used prescriptions for defining the initial vacuum
state -- the Bunch-Davies prescription and the adiabatic vacuum
prescription (even if the adiabaticity order to which the vacuum is
specified is infinity) -- fail to specify the vacuum uniquely in
generic inflationary spacetimes in which the total duration of
inflation is finite. This conclusion holds despite the absence of any
trans-Planckian effects or effective field theory cutoff related
effects.  We quantify the uncertainty which is applicable to slow roll
inflationary scenarios as well as for general FRW spacetimes and find
that the uncertainty is generically small. This uncertainty
should be treated as a minimal uncertainty that underlies all
curvature perturbation calculations. 
\end{quote}
\end{titlepage}

\baselineskip=24pt


\section{Introduction}
Recent measurements of the Cosmic Microwave Background (CMB)
 temperature fluctuations \cite{wmap} give strong
support for the picture that inflation gave rise to a flat spatial
geometry and the scale invariant energy density fluctuations on
superhorizon scales needed for large scale structure formation. As a
generic prediction of inflation, the primordial scale invariant energy
density fluctuations on superhorizon scales can be calculated
perturbatively once the model of inflation is specified, including the
prescription for the vacuum ({\it i.e.} Fock basis) of the inflaton
field. 

To see what one means by specifying a vacuum in the canonical
formalism, consider the simple case of a 
quantized scalar field $\phi$ in an FRW-type
universe with the metric of the form
\begin{equation}
ds^2 =a^2(\tau) (d\tau^2 - d\vec{x}^2)
\end{equation}
(conformal time coordinates) where we will restrict ourselves to flat
spatial sections for simplicity.  Without any nongravitational
interactions, the field in the Heisenberg representation has an
expansion
\label{h}
\begin{equation}
\phi({\bf x},\tau)=
\int \frac{d^3\!k}{(2 \pi)^{3/2} a(\tau)} \left[a_k h_k(\tau) 
e^{i
k \cdot {\bf x}} + a_k^\dagger h_k^*(\tau) 
e^{-i k \cdot {\bf x}} \right]
\label{eq:fundmodeexp}
\end{equation}
where the $a_k$ is the annihilation operator which annihilates the
vacuum and defines the Fock space.  Because the creation and
annihilation operators obey the commutator $[a_{k_1}, a_{{
k}_2}^\dagger] = \delta^{(3)} (k_1 -k_2)$, the $h_{ k}$'s obey a
normalization condition $h_k h_k^{'*} - h_k' h_k^* = i$ to satisfy the
canonical field commutators (henceforth, all primes or dots on
functions of $\tau$ refer to derivatives with respect to $\tau$ as
usual). Because the Heisenberg equation of motion for $\phi$ will
force $h_k(\tau)$ to obey a second order ordinary differential
equation, it has two independent solutions, and hence there needs to
be a boundary condition or prescription to specify $h_k$ thereby
defining the vacuum state.

Following the procedure in Minkowski space is one way of defining the
vacuum (i.e. specifying $h_k$): define $h_k$ to be the positive
frequency eigenvector of the time translational Killing
vector. Because cosmologically interesting spacetimes do not have
timelike Killing vectors, there is no a priori unique definition in
specifying the vacuum even without any non-gravitational
interactions. Indeed, even with a timelike Killing vector, as in de
Sitter space, there can be some ambiguity
\cite{chernikov,tagirov,mottola,allen}.  More generally, the simple
reason for a lack of a good definition of vacuum state is just a
problem of lacking asymptotically free states.  The classic studies of
quantum fields in curved spacetime (for good reviews, see for example
\cite{deWitt,fulling,birrelldavies,wald} and references therein) had
focused primarily on systematically identifying the ambiguities and
exploring various prescriptions that can remove them whenever
possible.

Two widely used prescriptions for defining vacuum are what is usually
called the Bunch-Davies prescription \cite{bunch} and the adiabatic
vacuum prescription of Parker and Fulling
\cite{parker69,parkerfulling,birrelldavies}.  The main point of this
paper is to show that both prescriptions, even when the adiabatic
order of the vacuum construction is {\em infinity} (i.e. the best one
can do), do not specify the vacuum uniquely in most situations.  This
is true as well in one of the most important case of generic
inflationary spacetimes\footnote{Here, the ``vacuum'' of the
inflationary era is referring to a no particle state of curvature
perturbations.} in which the total duration of inflation is finite,
despite the {\em absence} of any trans-Planckian effects or effective
field theory cutoff related effects \cite{transpl,danielsson}.  (For
arguments against the trans-Planckian and cutoff effects, see
\cite{Einhorn:2002nu,Bozza:2003pr,Starobinsky:2001kn,Starobinsky:2002rp}.)
The simple reason is that both methods rely on an asymptotic
definition of vacuum, leaving an ambiguity inherent in any asymptotic
expansion.  Note that the ambiguity that we are focusing on is also
independent of the effects due to transition {\em into} inflation
explored by Ref.~\cite{Grishchuk,turner,Burgess:2002ub}.  In addition
to showing the existence of an ambiguity, we estimate the uncertainty
and its implication for inflationary prediction of density
perturbations.

This work should complement the recent efforts
\cite{transpl,danielsson,Burgess:2002ub,Maldacena:2002vr,sr}
to uncover small quantum effects for the CMB measurements.  After all,
to see whether small effects can be measured, one must understand the
inherent theoretical uncertainty in the calculations.  Our uncertainty
should be viewed as a {\em minimal} uncertainty that underlies all of
these calculations, if one accepts either the adiabatic vacuum
formalism or the Bunch-Davies vacuum formalism.

In most generic situations, the uncertainty is extremely small.
Indeed, in view of our work, it should be clear that the vacuum used
in the work of Ref.~\cite{danielsson} does not constitute the best
adiabatic vacuum possible in a realistic inflationary scenario without
dS invariance.  From the adiabatic vacuum formalism, it should be
considered an ``excited'' state.  Even from a Hamiltonian minimization
point of view of Ref.~\cite{Bozza:2003pr}, one reaches a similar
conclusion.  To keep our presentation short as possible and to
emphasize its independence from trans-Planckian issues, we will not
discuss this point further in this paper.  However, the reader should
note that the existence of a cutoff does not change any of the results
in this paper for the adiabatic vacuum.

The order of presentation is as follows.  We start off by reviewing
the physical reasons for the ambiguities of a quantum vacuum in a
cosmological spacetime.  We then define and estimate the generic
uncertainties of the ``Bunch-Davies'' and the adiabatic vacuum.  The
following section gives an example of how nonperturbative quantities
can be calculated (despite the usual inherent uncertainties) in an
adiabatic vacuum in a very special limiting situation.  In
Sec.~\ref{sec:slowroll}, we compute what can be seen as a slightly
more precise uncertainty in the adiabatic vacuum during slow roll
inflationary spacetimes.  Finally, we summarize and conclude.

\section{Ambiguities of vacua}

From a traditional particle physicist's point of view, vacuum can be
defined as the state of no real particles.  As has been well studied
since 1960's (see for example \cite{deWitt,fulling,birrelldavies,wald}
and references therein), this notion of vacuum is well known to be
ambiguous: particles cannot always be unambiguously defined in the
presence of a background field.  For example, suppose one defines
particles as the eigenstates of the momentum operator.  These states
as wavefunctions must then necessarily be spacetime translation
eigenstates.  However, if there is no spacetime translational symmetry
of the background spacetime, there cannot be such an eigenstate.

Even if the spacetime were flat Minkowski space, if one were to define
particles empirically with an idealized detector, whether or not the
detector registers particles depends in general on the motion of the
detector.  If the detectors were restricted to geodesics of the
background spacetime, only in Minkowski spacetime, would all the
geodesic detectors agree to no particle detection \cite{birrelldavies}.

Because of this property of the Minkowski space, one may then try to
argue that it is best to abandon the notion of a vacuous curved
spacetime and treat it as a collection of gravitons in a Minkowski
background.  In this case, by definition the curved spacetime is not
vacuous, although the state with background gravitons may be
considered to be vacuous of some other field, say the inflaton field
$\phi$.  Unfortunately, partly due to $\phi$ interactions with the
background graviton fields, one is again forced back to having some
ambiguities in defining the vacuum for $\phi$.  Hence, seen this way,
we see the problem of the ambiguity of the vacuum, say with respect to
the field $\phi$, is not special to curved spacetime geometry, but to
any situation in which there is a background field that interacts with
the field $\phi$.  Just as in any other interacting quantum field
theory, one may try to define free asymptotic states for $\phi$ and
the graviton, and treat the interactions perturbatively.  This would
work if there is an asymptotically flat region of spacetime.
Unfortunately, the key difference in the curved background situations
of interest to inflationary cosmology is that there are no such
asymptotically flat regions.\footnote{There may be a way of obtaining
a physically sensible answer by artificially turning on and off the
gravitation, but we will not pursue this line of reasoning in this
work.}  Hence, the problem of vacuum prescription can be rephrased as
trying to define a vacuum state in the absence of free asymptotic
states.

Note that instead of relying on a particle definition of vacuum which
is inherently nonlocal and observer dependent (the number of particles
detected by any physical detector can be zero in one frame but not in
another, even in Minkowski space), one can characterize the vacuum in
terms of vacuum expectation value of the stress tensor, which has the
advantage of being covariant (if it is zero in one frame, it is zero
in all frames of reference).  However, even here, the stress tensor
vacuum expectation value is sensitive to the ambiguities in the
boundary conditions and renormalization prescriptions of the
correlation functions.  These boundary condition sensitivities then
reflect the ambiguities of the vacuum.

\section{The ``Bunch-Davies'' Vacuum and the Adiabatic Vacuum}
In this section, we would like to explain the two commonly used
prescriptions that will be the focus of this paper.

\subsection{``Bunch-Davies'' vacuum}
In the FRW cosmological context, the most well known and appealing
prescription for identifying the vacuum is what is commonly called the
Bunch-Davies prescription, which states that the positive frequency
mode function $h_k$ (see Eq.~(\ref{eq:fundmodeexp})) should match
asymptotically to the Minkowski space prescription in the limit that
the physical momentum \( k/a \) is much larger than the background
geometry curvature scale (\( H=\dot{a}(\tau)/a^2 \) ):
\begin{equation}
\lim_{ k/(aH)\rightarrow \infty} h_k \sim \frac{1}{\sqrt{2 k}} e^{-ik \tau}. 
\end{equation}
In some cases, this uniquely specifies the vacuum.

However,when this prescription is carried out at a fixed cosmological
time (i.e. \( k\rightarrow \infty \) with \( a(t) \) fixed), then the
vacuum is not unique even in the simplest situations. Since inflation
generally did not last an infinitely long time, the Bunch-Davies
prescription cannot be applied in the asymptotic past limit, and the
ambiguities associated with \( k\rightarrow \infty \) are
nonvanishing.  This is independent of the existence of cutoffs or any
trans-Planckian physics.

For example, consider a massless scalar field minimally coupled to
Einstein gravity in a patch of de Sitter space
\begin{equation}
S=\int d^4\!x\, \frac{a^2}{2}\left( \dot{\phi}^2 - (\nabla
\phi)^2 \right).
\label{eq:masslessscalar}
\end{equation}
where $a(\tau)=-1/(\tau H)>0$ with $1/H >0$ being the dS radius. 
The resulting mode equation is
\begin{equation}
\ddot{h}_k(\tau) + w_k^2(\tau) h_k(\tau) = 0,
\label{modeequation}
\end{equation}
where dots stand for derivatives with respect to the conformal time $\tau$
and 
\begin{equation}
w_k^2= k^2 -  \frac{\ddot{a}}{a}=k^2-\frac{2}{\tau^2} \ .
\label{eq:frequency}
\end{equation}
The general positive frequency mode solution is
\begin{equation}
\label{eq:masslessminimal}
h_{k}=A_{k}\frac{e^{-ik\tau
}}{\sqrt{2k}}\left(1+\frac{iH}{(k/a)}\right)+B_{k}\frac{e^{ik \tau
}}{\sqrt{2k}}\left(1-\frac{iH}{(k/a)}\right)
\end{equation}
 where \( A_{k} \) and \( B_{k} \) satisfies\begin{equation}
|A_{k}|^{2}-|B_{k}|^{2}=1\end{equation} 
due to the normalization conditions given below Eq.~(\ref{eq:fundmodeexp}).
At any given time \( \tau  \), one can impose
that the modes become Minkowskian as \( |(k/a)/H|\rightarrow \infty
\).
Explicitly, this amounts to a functional matching of the form\begin{equation}
h_{k}\sqrt{2k}e^{ik\tau }\rightarrow 1\end{equation}
for  as \( |k\tau |\rightarrow \infty  \) for all \( \tau <0
\).
Now, note that since \( \tau \rightarrow -\infty  \) in
Eq.~(\ref{eq:masslessminimal})
gives\begin{equation}
h_{k}\rightarrow A_{k}\frac{e^{-ik\tau
}}{\sqrt{2k}}+B_{k}\frac{e^{ik\tau }}{\sqrt{2k}}\end{equation}
 for all \( k>0 \), this \textbf{uniquely} specifies \( A_{k}=1 \)
and \( B_{k}=0 \) for all \( k>0 \).

However, consider the situation in which we make the restriction \(
|\tau |<|\tau _{0}| \) where \( \tau _{0} \) is the time at the
``beginning'' of inflation.  Now, the limit \( |(k/a)/H|\rightarrow
\infty \) can be taken only by taking \( k\rightarrow \infty \) with
\( a>a(\tau _{0}) \).  This means that we only know, for example, that
\begin{equation} 
B_{k}\rightarrow {\cal O}(1/k^n)
\end{equation}
with $n>0$ as \( k\rightarrow \infty \).\footnote{Note that $\exp{(-
ik\tau)}$ does not have an asymptotic expansion in real $k$ as
$k\rightarrow \infty$ because of an essential
singularity.} Therefore, if we just impose \( k\rightarrow \infty \)
with a bounded \( a(\tau ) \), the Bunch-Davies prescription does
\textbf{not} uniquely specify a vacuum. Applied to the case of
Eq.~(\ref{eq:masslessminimal}), one must allow the ambiguity
\begin{equation} 
B_{k}\la {\cal O}\left(\frac{H}{k/a(\tau_{0})}\right)\, .
\label{eq:bunchdaviesuncertainty}
\end{equation}
A useful quantity to characterize the properties of the quantum
perturbations of a massless scalar field during inflation is the power
spectrum.  For a generic quantity $g({\bf x},\tau)$, which can
expanded in Fourier space as 
\begin{equation}
 g({\bf x},\tau)=\int\,\frac{d^3 k}{(2\pi)^{3/2}}\,
e^{ik x}\, g_{k}(\tau), 
\end{equation}
the power spectrum can be defined as 
\begin{equation}
 \langle 0|g^{*}_{k_1}g_{k_2}|0\rangle \equiv\delta^{(3)}\left(k_1-
k_2\right)\,\frac{2\pi^2}{k^3}\, {\cal P}_{g}(k), 
\end{equation} 
where $\left|0\right.\rangle$ is the vacuum quantum state of the
system. This definition leads to the usual relation \be \langle
0|g^2({\bf x},t)|0\rangle=\int\,\frac{dk}{k}\, {\cal P}_{g}(k)\, .  \ee
If we compute the variance of the perturbations of the  $\phi$
field
\begin{eqnarray}
\langle 0|\left(\phi({\bf x},\tau)\right)^2|0\rangle&=&
\int\,\frac{d^3k}{(2\pi)^3}\,\left|\phi_k\right|^2\nonumber\\
&=&\int\,\frac{dk}{k}\,\frac{k^3}{2\pi^2 a^2}
\,\left|h_k\right|^2\nonumber\\
&=& \int\,\frac{dk}{k}\,{\cal P}_{\phi}(k)\, ,
\end{eqnarray}
we may infer  the power spectrum of the
fluctuations of the scalar field $\phi$ to be 
\begin{equation}
{\cal P}_{\phi}(k)\equiv\frac{k^3}{2\pi^2}
\,\left|\phi_k\right|^2.
\label{spectrum}
\end{equation} 
Therefore, for a massless scalar
field in de Sitter space, we obtain on superhorizon scales the power spectrum 
\begin{equation}
{\cal P}_{\phi}(k)=\left(\frac{H}{2\pi}\right)^2\left| A_k-B_k\right|^2 \left(\frac{k}{aH}\right)^{n_\phi-1}\, ,
\label{fff}
\end{equation}
with $n_\phi= 1$. From Eq.~(\ref{fff}) 
we infer then that any ambiguity in the
parameter $B_k$ implies an ambiguity in the power spectrum ${\cal P}_{\phi}$
of the form
\begin{equation}
\left|\frac{\delta{\cal P}_{\phi}(k)}{{\cal P}_{\phi}(k)}\right|\simeq
2\, \left| {\rm Re} \, B_k \right| \simeq {\cal O}(e^{-(N_0-N_k)})\, ,
\label{eq:bunchdaviesspecunc}
\end{equation}
where $N_k$ denotes the number of $e$-foldings before the end of
inflation when a given wavelength $\lambda=a/k$ leaves the horizon
during the de Sitter stage and $N_{\rm 0}$ denotes the number of
e-foldings before the end of inflation when the spacetime can be
considered a vacuum (which has an upper bound of total number of
e-foldings for inflation).  This power spectrum can be seen as an
approximation to the curvature perturbation power spectrum, and the
vacuum here can be seen as the vacuum with respect to curvature
perturbations. Length scales of interest for the the CMB anisotropies
give $N_k$ of order of 60, and therefore one expects the theoretical
ambiguity on the power spectrum to be sizeable if the total duration
of the de Sitter stage corresponds to a number of $e$-foldings not far
from 60.  Of course, in this case it would be difficult to assume that
the spacetime is in a vacuum state (see for example
\cite{Grishchuk,turner,Burgess:2002ub}).

\subsection{Adiabatic vacuum \label{adiabaticvacuumsection}}
In situations in which the vacuum is defined through the notion of
particles, the natural fundamental operator is the number operator
which counts the number of particles with momentum $k$.  Parker in
Ref.~\cite{parker69} postulated certain reasonable conditions that the
number operator for a scalar field must satisfy in a FRW universe.
The conditions were as follows:
\begin{enumerate}
\item $N_k$ be Hermitian due to the counting interpretation.
\item When the expansion is stopped at any time (i.e. $\dot{a}/a=0 $),
the operator becomes the usual Minkowski number operator.
\item The vacuum expectation value of the number operator varies
slowly as possible with time as the expansion rate $\dot{a}/a$ becomes
arbitrarily slow.
\end{enumerate}
The first two conditions are obviously reasonable, and give rise to
the definition of the number operator as 
\begin{equation}
N_k(\tau_1) \equiv a_k^{(\tau_1) \dagger} a_k^{(\tau_1)}
\end{equation}  
where the superscripted $\tau_1$ refers to the time at which the
boundary condition for vacuum is set (or equivalently, the boundary
conditions for $h_k(\tau)$).  

The third condition is the statement that the vacuum should be defined as to
keep the number of particles as unchanging as possible.  This third
condition is what Parker has called the {\em minimization postulate}
\cite{parker69} and was later developed further by Parker and Fulling
\cite{parkerfulling}.  The formalism developed based on satisfying these
conditions is called the adiabatic vacuum formalism
\cite{parkerfulling,birrelldavies}.  

It is important to note that to define the number operator $N_k$, one
must define what one means by a particle at any given time, $\tau_1$.
On the other hand, to have defined a vacuum at an earlier time
$\tau_0<\tau_1$ (recall that Heisenberg representation states such as
the vacuum are time independent), one needed a definition of a
particle at time $\tau_0$.  Hence, the present formulation requires
that one define particles at two different times.  This should be
contrasted with the formalism of computing the vacuum expectation
value of the stress tensor, where the definition of particles (or
equivalently the vacuum) needs to be defined only once, since the
stress energy tensor is a local quantity which does not rely on the
basis of Fourier expansion.  However, since we will be taking the same
prescription for particles at time $\tau_0$ and time $\tau_1$ in such
a way that the total number of conditions that has to be specified is
the same as in the case of computing the vacuum expectation value of
the stress tensor, this inherently nonlocal definition of vacuum will
probably also minimize the growth of the stress tensor to a large
extent.  Nonetheless, it is not obvious how the conclusions of our
analyses would differ if stress tensor vacuum expectation values are
used instead of $N_k$ in carrying out the minimization postulate.  We
will defer this question to a future work and focus on the particle
based formalisms in this paper.

The adiabatic formalism specifies the value and the first time
derivative of the mode function $h_k$ at a \emph{fixed} time for
\emph{any} fixed momentum. The boundary condition data is specified by
matching to an asymptotic expansion (adiabatic expansion) of the mode
equation to {\em all} orders in the asymptotic parameter (even though
the expansion does not converge in general, by the very nature of an
asymptotic expansion).  In de Sitter space, for example, the infinite
adiabatic order vacuum is identical (up to the uncertainties inherent
to the adiabatic vacuum) to the Bunch-Davies vacuum.  In inflationary
spacetimes in which the scale factor cannot be made arbitrarily small
due to the finite duration of the inflationary phase or in which the
momentum cannot be made arbitrarily large due to a cutoff, the
adiabatic formalism naively has a chance to still specify a unique
vacuum, unlike the Bunch Davies prescription.  However, we will find
that the adiabatic prescription suffers from an ambiguity problem as
well.

The construction of the adiabatic vacuum is as follows.  First, define
the concept of an adiabatic order as the power of $1/T$ that results
for any term in an $1/T \rightarrow 0$ asymptotic expansion after one
makes the transformation $\tau \rightarrow \tau$ and $d/d\tau
\rightarrow T^{-1} d/d\tau$ in the differential equation for the modes
$h_k$.\footnote{Note that the point about only the derivative being
transformed has been missed by a recent paper \cite{sr}.}. Then, any prototypical mode equation of the form
\begin{equation}
\ddot{h}_k(\tau)+ w_k^2(\tau) h_k(\tau)=0 
\label{eq:genericmode}
\end{equation}
with $w_k^2= k^2+ m^2 a^2(\tau)+(6 \xi -1) \ddot{a}/a$ (where $\xi$ is
a constant and $m$ is the mass of $\phi$) turns into
\begin{equation}
\frac{1}{T^2}\ddot{\widetilde{h}}_k(\tau)+ \widetilde{w}_k^2 \widetilde{h}_k(\tau)=0
\label{eq:wkbmodeequation}
\end{equation}
where $\widetilde{w}_k^2 \equiv k^2+ m^2 a^2(\tau)+(6 \xi -1) \ddot{a}/(T^2
a)$ and a tilde has been added to $h_k$ to be a reminder that this
function carries a fictitious parameter $T$ (later this will be set to
1 at which time the function will be denoted as $h_k$).  Now, make a
change in variables from $\widetilde{h}_k$ to $W_k$ by writing
\begin{equation}
\widetilde{h}_k=\frac{1}{\sqrt{2 W_k}} e^{\left(-i \int^\tau W_k(\tau')
d\tau' T\right)}
\label{eq:newsol}
\end{equation}
(where the $T$ in the exponent should be noted) and from Eq.~(\ref{eq:wkbmodeequation}), we obtain a new differential equation
\begin{equation}
W_k^2 = \widetilde{w}_k^2 - \frac{1}{2 T^2} \left[\frac{\ddot{W}_k}{W_k} - \frac{3}{2}
\left(\frac{\dot{W}_k}{W_k}\right)^2\right]\, .
\label{eq:wkbiter}
\end{equation}
We can then define a map
\begin{equation}
A\left[W_k^{[n]}\right]= 
\sqrt{ \widetilde{w}_k^2 - \frac{1}{2T^2} \left[ \frac{\ddot{W}_k^{[n]}}{W_k^{[n]}} - \frac{3}{2}
\left(\frac{\dot{W}_k^{[n]}}{W_k^{[n]}}\right)^2\right] }
\label{coolmap}
\end{equation}
which is a map that raises the adiabatic order by two and also 
define
\begin{equation}
W_k^{[n+2]}= A\left[W_k^{[n]}\right],
\label{eq:recursion}
\end{equation}
where the superscript denotes the adiabatic order and
$W_k^{[0]} \equiv  \sqrt{ k^2 + m^2 a^2} $.  All of this construction
lead to an approximate mode equation solution
good to $A$th adiabatic order in asymptotic expansion in $1/T
\rightarrow 0$ of
\begin{equation}
h_k^{[A]} = \frac{1}{\sqrt{2 W_k^{[A]}}} e^{\left(-i \int^\tau
W_k^{(A)}(\tau') d\tau' T\right)}\ .
\label{template}
\end{equation}
This asymptotic expansion solution will serve as a template, just to
set the boundary conditions for the mode function $\widetilde{h}_k$ which
can in general be written as
\begin{equation}
\widetilde{h}_k(\tau) = A_k f_k(\tau) + B_k f_k^*(\tau)
\label{eq:vacuumparam}
\end{equation}
where $f_k$ are exact basis functions satisfying
\eqr{eq:wkbmodeequation}.  Specifically, we define the $A$th adiabatic
(order) vacuum at time $\tau_0$ by using the boundary condition
\begin{eqnarray}
\widetilde{h}_k^{\tau_0}(\tau_0) &=& h_k^{[A]}(\tau_0) + 
{\cal O}(1/T^{(A+1)})\nonumber\, ,\\
\dot{\widetilde{h}_k}^{\tau_0}(\tau_0) &=& \dot{h}_k^{[A] }(\tau_0) + 
{\cal O}(1/T^{(A+1)})\, ,
\label{matching}
\end{eqnarray}
where the left-hand side is the {\em exact} mode solution to the
prototypical differential equation Eq.~(\ref{eq:wkbmodeequation}) and
the boundary conditions are enforced to only order $1/T^{(A+1)}$ as $T
\rightarrow \infty$.\footnote{The correction is order $1/T^{(A+1)}$
and not of order $1/T^{(A+2)}$ because there is an integration in the
exponent.} After the boundary conditions are set, one sets the vacuum
mode of \eqr{eq:fundmodeexp} as
\begin{equation}
h_k^{\tau_0}(\tau) \equiv \widetilde{h}^{\tau_0}_k(\tau)|_{T=1}
\label{eq:removeT}
\end{equation}
which removes the fictitious parameter $T$.

To see that this construction satisfies the minimization postulate if
we take the vacuum adiabatic order $A$ to $\infty$, let us compute the
vacuum expectation value of the number operator corresponding to the
number density per mode.  This requires a Bogoliubov transformation
from the vacuum mode solution with the boundary condition at
$\tau=\tau_0$ into the one with the boundary condition at $\tau=
\tau_1$ (any later time at which the particles are no longer being
created).  Defining the Bogoliubov transformation as
\begin{equation}
h_k^{\tau_1}(\tau)= \alpha_ kh_k^{\tau_0}(\tau) +
\beta_k h_k^{* \tau_0}(\tau)
\label{eq:bogoliubovdef}
\end{equation} the number density of particles per momentum $k$
is given by $n_k=\left|\beta_k\right|^2$.  If the vacuum
in the past is defined at $\tau=\tau_0$ with infinite adiabatic order
boundary condition and the vacuum today is defined at $\tau=\tau_1$
with infinite adiabatic order boundary condition, carrying out the
Bogoliubov transformation with the solution written in the form
Eq.~(\ref{eq:newsol}), one finds

\begin{equation} 
_{\tau_0}\langle0| N_k(\tau_1)|0 \rangle_{\tau_0} 
\propto | \beta_k(\tau_1, \tau_0)|^2 = \frac{1}{4
W_k^{\tau_0} W_k^{\tau_1}} \left\{ \frac{1}{4} \left( \frac{\dot{W}_k^{
\tau_0}}{W_k^{\tau_0}} - \frac{\dot{W}_k^{\tau_1}}{W_k^{\tau_1}} \right)^2
+ ( W_k^{\tau_0}- W_k^{\tau_1})^2 \right\}
\label{eq:numdensasymp}
\end{equation}
where the right hand side (composed of exact solutions to
Eq.~(\ref{eq:wkbiter})) can be evaluated at any $\tau$ and the
superscripts indicate the time at which the boundary conditions were
placed.

Now, the minimization postulate is satisfied if $ \frac{d^n
}{dt^n}(_{\tau_0}\langle0| N_k(\tau_1)|0 \rangle_{\tau_0})$ is
minimized for all non-negative integers $n$ if the expansion rate can be
turned off arbitrarily slowly.  For any $a(\tau)$, we can affect the
slowly turning off of the expansion rate by reintroducing the
adiabatic order parameter $1/T \rightarrow 0$ as before.  Since by
construction, the $W_k^{\tau_i}$ functions in
Eq.~(\ref{eq:numdensasymp}) when expanded in $1/T$ match the
asymptotic expansions in $1/T$ of $W_k^{[\infty]}$ of
\eqr{eq:recursion}, as long as the asymptotic expansion is uniform in
$\tau$ between $\tau_0$ and $\tau_1$ (no singularities in the
asymptotic expansion occur between $\tau_0$ and $\tau_1$), the right
hand side of Eq.~(\ref{eq:numdensasymp}) vanishes identically when
expanded in $1/T$. Hence $\frac{d^n}{dt^n}(_{\tau_0}\langle0|
N_k(\tau_1)|0 \rangle_{\tau_0}) $ will fall off faster than any finite
power of $1/T$ as $T\rightarrow \infty$, thus satisfying the
minimization condition.

In general, one must remember that $W^{[\infty]}$ does not exist
because the $W^{[A]}$ construction procedure generates an asymptotic
expansion about a nonanalytic point rather than a convergent
series. In other words, for a fixed order $A$ (and with $T=1$), there
exists only an $A$-dependent region in time for which $W^{[A]}$
approximates well the exact solution to Eq.~(\ref{modeequation}), with
the leading error on the approximation growing with $A$ for any fixed
time $\tau$. Hence, when the limit $A\rightarrow \infty$ is taken
first, the time region in which the approximation is valid can shrink
to 0 for a fixed $T=1$.

As an example of an adiabatic vacuum, let us apply this formalism to
the case of massless scalar field in dS space (considered in
Eq.~(\ref{eq:masslessscalar})).  The iteration map Eq.~(\ref{coolmap})
produces up to sixth order
\begin{eqnarray}
W_k^{[0]\ 2} & = & k^2 \nonumber\\
W_k^{[2]\ 2}&=&k^2\left[ 1-\frac{2}{(T k\tau)^2}  \right] \nonumber
\\
W_k^{[4]\ 2}&=&k^2\left[1-\frac{2}{(T k\tau)^2}+\frac{3}{(T k\tau)^4}
+{\cal O}\left(\frac{1}{(T k\tau)^6}\right)  \right] \nonumber\\
W_k^{[6]\ 2}&=&k^2\left[1-\frac{2}{(T k\tau)^2}+\frac{3}{(T k\tau)^4}
-\frac{4}{(T k\tau)^6}+{\cal O}\left(\frac{1}{(T k\tau)^8}\right) \right].
\end{eqnarray}
The template asymptotic expansion $\widetilde{h}_k^{[A]}$ obtained
from $W^{[A]}$ approximates the exact solution with an error of order
$1/T^{A+1}$.  For example, one can write down explicitly using
$W_k^{[4]}$ (dropping the higher order corrections to it)
\begin{equation}
\widetilde{h}_k^{[4]}=\frac{e^{-i k \tau T}}{\sqrt{2 k}}\left[1 
- \frac{i}{k \tau T} +
\frac{2 i}{5 k^5 \tau^5 T^5} + {\cal O}(1/T^6)\right]
\end{equation}
where the $1/T^5$ term is the leading uncertain term (displayed just
for clarity), and up to $1/T^4$, the template function matches the
first term of the exact solution
\begin{equation}
\widetilde{h}_{k}=A_{k}\frac{e^{-ik \tau T
}}{\sqrt{2k}}\left(1-\frac{i}{(k \tau T)}\right)+B_{k}\frac{e^{ik \tau T
}}{\sqrt{2k}}\left(1+\frac{i}{(k \tau T)}\right)\, .
\label{eq:exactsolwT}
\end{equation}
Carrying out the matching procedure of Eq.~(\ref{matching}) except up
to 5th adiabatic order instead of the just the required 4th, one finds
\begin{eqnarray}
A_k& = & 1  + \frac{3 i}{2 k^5 \tau_0^5 T^5} + ... \\
B_k & = & 0  + {\cal O} (1/T^6)
\end{eqnarray} 
where we have displayed one of the uncertain terms explicitly for
clarity.  Note that the coefficients in general depend on $\tau_0$
(when the boundary condition was placed), but the sensitivity to it is to
higher adiabatic order. In general, the coefficients $A_k$ and
$B_k$ will be of the form
\begin{eqnarray}
A_k &=& 1+ {\cal O}(1/T^{A+1})\, ,\nonumber \\
B_k &=& 0+ {\cal O}(1/T^{A+1})
\end{eqnarray}
for an $A$th order adiabatic vacuum.  In the limit that $A\rightarrow
\infty$, the uncertainty drops off faster than any finite power of $1/T$.

Recall that in the Bunch-Davies prescription, one could not remove the
uncertainty in the choice of $\{A_k, B_k\}$ if we did not have information
to let $a(\tau)\rightarrow 0$, say because of the finite period of
inflation.  Since in the adiabatic prescription, the boundary
conditions may be set to infinite adiabatic order at a finite time,
let us see whether at finite initial time $\tau_0$ it is possible to
define an infinite adiabatic vacuum without any ambiguity.  By
inspection, we can cast $W^{[A] \ 2}$ in the form
\begin{equation}
W^{[A] \ 2}= k^2 \, \sum_{n=0}^{A/2} \,(-1)^n \,(n+1)
\left(\frac{1}{(T k \tau)^2}\right)^{n} + {\cal O}
\left(\frac{1}{T^{A+2}}\right)\, .
\end{equation}
This means that in the limit $A\rightarrow \infty$, $W^{[\infty] \ 2}$
 is expressed as
\begin{equation}
W^{[\infty] \ 2}= k^2 \, \sum_{n=0}^{\infty}\, (-1)^n\, (n+1)
\left(\frac{1}{k\tau T}\right)^{2n}
\end{equation}
which converges  to a simple function\footnote{Notice that
in the limit $\tau=\tau_0\rightarrow -\infty$ one recovers
$W^{[\infty] \ 2}= k^2$, {\it i.e.}
the infinite adiabatic order vacuum reduces to the BD vacuum.}
\begin{equation}
W^{[\infty] \ 2}= k^2 \, \frac{(k\tau T)^4}{(1+(k\tau T)^2)^2}.
\end{equation}
Miraculously, the series converges in this simple
situation!\footnote{We will later give an example where there is no
convergence.}  Indeed, inserting $W^{[\infty]}$ into (\ref{template}),
we obtain at any time $\tau$
\begin{equation}
h^{[\infty]}_k=
\frac{e^{-ik\tau T}}{\sqrt{2k}}  
\left(1-\frac{i}{k\tau T}\right),
\label{eq:exactwsummed}
\end{equation}
which in view of \eqr{matching} will set $A_k=1$ and $B_k=0$ in
\eqr{eq:exactsolwT} (note that we have normalized the function
properly with the arbitrary integration constant freedom).

This seems to indicate that the vacuum has been uniquely specified.
Unfortunately, this is not true since if an infinite adiabatic
order vacuum with \( T\rightarrow \infty \) is chosen to be \( \widetilde{h}_k(\tau
) \), we can always choose another vacuum
\begin{equation}
{\underline {\widetilde h}}_k=\sqrt{1+|B_{k}|^{2}}{\widetilde h}_k(\tau )
+B_{k}{\widetilde h}_k^{*}(\tau )
\end{equation} if \( B_{k} {\widetilde h}_k^*(\tau ) \) falls off
faster than any finite power of \( 1/T \). The infinite order
adiabatic vacuum boundary conditions do not distinguish \( {\widetilde
h}_k \) and \( {\underline {\widetilde h}}_k \).  Indeed, this can be seen directly
in \eqr{matching} because the ``equation'' is asymptotic, up to terms
that vanish faster than $A$th power of $1/T$, where in the infinite
adiabatic order case $A=\infty$: i.e. we should have written instead
of \eqr{eq:exactwsummed}, the equation
\begin{equation}
h^{[\infty]}_k=
\frac{e^{-ik\tau T}}{\sqrt{2k}}  
\left(1-\frac{i}{k\tau T}\right) + {\cal O}(1/T^\infty)
\end{equation}  
where $ {\cal O}(1/T^\infty)$ indicates the nonperturbative term
possibly dropped in the matching to the template function. Since $T$
follows every factor of $k$, we can guess the uncertainty by assuming
an exponential function for $B_k$ as
\begin{equation}
B_k \approx \exp(-k T/(a(\tau_0) H(\tau_0)))
\label{eq:bkexp}
\end{equation}
which should be compared to \eqr{eq:bunchdaviesuncertainty}.  (Note
that this would also serve as a good estimate of uncertainties even
outside of inflationary phase.)  This leads to
\begin{equation}
\left|\frac{\delta{\cal P}_{\phi}(k)}{{\cal P}_{\phi}(k)}\right|
\simeq {\cal O}(\exp(- e^{N_{\rm 0}-N_k}))\, ,
\label{eq:bkexpinton}
\end{equation}
where we have used the same notation as \eqr{eq:bunchdaviesspecunc}.
Because of the double exponential, the uncertainty is very quickly
negligible as $N_0$ becomes larger than $N_k$.

In the case in which the expansion in $1/T$ stops converging, as in the
 quasi-dS case, we may be able to estimate the uncertainty in the 
adiabatic vacuum
formalism a little less arbitrarily (compared to \eqr{eq:bkexp}) as
follows.  \label{stopsconverging} Indeed, if at
some adiabatic order $n_*$ the expansion in $1/T$ stops converging (with $T=1$, the higher order terms give a larger correction to the
mode solution than the $n_*$th order term), this signifies that the
nonadiabaticity in the system enters at the $n_*$th derivative.
Therefore, as far as assigning an uncertainty to the adiabatic vacuum
is concerned, instead of using \eqr{eq:bkexp}, we may estimate the
uncertainty in fixing the vacuum to simply be the $n_*$ term in the
asymptotic expansion.  Although the uncertainty obtained this way is
not ``nonperturbative'', it is not clear what the physical advantage
is in artificially ``turning off'' the expansion (using $1/T$) beyond
the $n_*$ derivative which is in some sense the limit of adiabaticity
characteristic of the physical system.  We will later use this
approach to compute the uncertainty for the slow roll inflationary
case.

There is an exception to the existence of an uncertainty in the
adiabatic vacuum formalism, which is already suggested by the estimate
in \eqr{eq:bkexp} in the case that $a(\tau_0)=0$.\footnote{Here, we
are not claiming $a(\tau_0)=0$ generically defines an adiabatic point,
but merely that \eqr{eq:bkexp} suggests there are special points in
spacetime where the uncertainty vanishes.} Namely, if the boundary
conditions are set at a time in which the nonadiabaticity is
identically 0, then there is no expansion in $1/T$, and hence there is
no ambiguity in the adiabatic vacuum.  Indeed, the loss of time
translational invariance which is operationally at the heart of the
uncertainty of the vacuum disappears at the time when the
nonadiabaticity is identically 0.  The test of the disappearance of
the nonadiabaticity is that $W^{[A]}$ for any $A$ becomes identical,
or more explicitly
\begin{equation}
W^{[A]}(\tau_0)= W^{[0]}(\tau_0)
\label{eq:exceptionalcasetest}
\end{equation}
for all $A$ where $\tau_0$ is the time at which the vacuum is defined.
This time $\tau_0$ typically is $\pm \infty$, which means that an
asymptotic expansion in $\tau$ must be taken to match the boundary
conditions.  Hence, the reason why the ambiguity disappears in this
exceptional case can also be seen as due to the asymptotic expansion
being in time $\tau$ instead of $1/T$, since the coefficients in
\eqr{eq:vacuumparam} are time independent although they can be be $T$
dependent.

This exceptional case can be viewed as a meeting point of the
adiabatic vacuum formalism and the Bunch-Davies formalism.  Its spirit
is similar to the Bunch-Davies formalism in that the asymptotic
expansion can be in time $\tau$ instead of a fictitious adiabatic
parameter $T$. However instead of matching on to a Minkowski
prescription, it matches on to a zeroth adiabatic order WKB
prescription as in the adiabatic vacuum formalism.  Essentially, the
adiabatic formalism states that although not in general, sometimes a
unique vacuum can be defined in regions of spacetime where the
frequency is adiabatically a constant to an arbitrarily good degree
just as in the Bunch-Davies prescription, except with the frequency
given by the leading order WKB ansatz instead of a Minkowski
prescription (Minkowski prescription means that with the scale factor
frozen at a fixed value).

The reader should be aware that the adiabatic formalism is compatible
with the Bunch-Davies formalism.  For example, this formalism always
gives the same vacuum as the Bunch-Davies formalism whenever the
Bunch-Davies formalism can define a unique vacuum.  Also, we would
like to comment that although we would like to extend the
``uniqueness'' to the case in which WKB iteration \eqr{coolmap}
converges for $n=\infty$ (i.e. $W^{[\infty]}(\tau)$ is well defined
with $T$ fixed at a finite value), it is not obvious to us whether
there is any reason to consider such situations in any special manner
since the convergence of WKB iteration is not identical to the absence
of nonadiabaticity.  It merely means that the asymptotic expansion in
$1/T$ is being taken about an analytic point in the function
generating the asymptotic expansion.

\section{An Example of Calculable Nonperturbative Particle Production}

In this section, we will discuss a model from
Ref.~\cite{audretschschaefer} which illustrates the adiabatic
formalism in situations where nonperturbative particle production can
be unambiguously calculated.  Owing to the supposed background metric,
the vacuum construction will be unique, and hence exponentially
suppressed particle production will be calculable.  

Before moving on to the discussion, we would like to alert the reader
here that this model is also reviewed on pg.~70 of
Ref.~\cite{birrelldavies}.  However, their usage of adiabatic
prescription can be misleading (it was to us, at least).  As a
consequence of their loose usage, they give the wrong impression that
they can calculate reliably an exponentially suppressed particle
production even without taking to $ \pm \infty$ the time at which the
vacuum is specified.  For example, one sees that the matching in their
equation (3.117) involves taking $\lambda \rightarrow \infty$, which
by itself results in a possibly exponentially suppressed ambiguity in
the vacuum that we discussed in \eqr{eq:bkexp}.  However, as long as
the vacuum is specified at $\tau_0=-\infty$, we agree with their
results modulo their inconsequential phase error in their equation
(3.124).  We present our reanalysis here such that the reader would
not be confused about the discrepancy of how Ref.~\cite{birrelldavies}
seems to claim that nonperturbative calculations can be unambiguously
done in the context of the adiabatic vacuum formalism even without
taking $\tau_0 \rightarrow -\infty$ while we claim that such
calculations cannot be done, strictly speaking, without being able to
take such a limit for the time at which the vacuum is specified.

To see how the nonperturbative Bogoliubov coefficient arises in an
exactly solvable model of Ref.~\cite{audretschschaefer}, consider the
conformally coupled mode equation (i.e. \eqr{eq:genericmode} with
$\xi=1/6$) where the scale factor is given by\begin{equation}
\label{eq:symmetricscale}
a^{2}(\tau )=\alpha ^{2}+\beta ^{2}\tau ^{2}
\end{equation}
leading to  the effective frequency
\begin{equation}
w_{k}^{2}=m\beta \lambda +m^{2}\beta ^{2}\tau ^{2}
\end{equation}
 where\begin{equation}
\lambda \equiv \frac{m\alpha ^{2}}{\beta }+\frac{k^{2}}{m\beta }\, .
\end{equation}
Since we would like to carry out the adiabatic procedure, we scale the
derivatives w.r.t. \( \tau \) by \( 1/T \):\begin{equation}
\label{eq:modeqorig}
\left(\frac{1}{T^{2}}\partial _{\tau }^{2}+w_{k}^{2}\right){\widetilde{h}}_{k}=0 \, .
\label{eq:bouncemodeeq}
\end{equation}
For differential equations of the form \begin{equation}
\frac{d^{2}u}{dx^{2}}+(x^{2}+\widetilde{\lambda })u=0\, ,
\end{equation}
 the general solution is the parabolic cylinder function\begin{equation}
u=D_{-(1+i\widetilde{\lambda })/2}[\pm (1+i)x]
\end{equation}
and its complex conjugate where for our case, one should identify
\begin{equation}
x=\sqrt{m\beta T}\tau\, ,
\end{equation}
\begin{equation}
\widetilde{\lambda }=T\lambda\, ,
\end{equation}
where one notes that as \( T\rightarrow \infty  \), the solution
is effectively doubly scaled as \( \sqrt{T}\tau  \) and \( T\lambda  \).
Note also that the parabolic cylinder function \( D_{p}(z) \) is
defined by its boundary condition\begin{equation}
D_{p}(z)\sim z^{p}e^{-z^{2}/4}
\end{equation}
 as \( z\rightarrow \infty  \). However, this asymptotic expansion
is not the same expansion as the \( T\rightarrow \infty  \) expansion
since there is an effective double scaling of \( \tau  \) and \( \lambda  \).

Let us first see what the adiabatic vacuum procedure gives for the
particle production.  We begin by writing explicitly the exact mode
function (solution to \eqr{eq:bouncemodeeq}) as
\begin{equation}
\widetilde{h}_{k}=A_{k}f+B_{k}f^{*}
\end{equation}
 where\begin{equation}
\label{eq:parabolicmodebasis}
f(\tau)=\frac{T^{1/4}}{(2m\beta )^{1/4}}e^{-\pi \lambda T/8}D_{-(1-i\lambda T)/2}[(i-1)\sqrt{m\beta T}\tau ]
\label{eq:parabolmode}
\end{equation}
which as one sees has the curious double scaling of \( \tau \) and \(
\lambda \) mentioned before. We need to obtain an asymptotic expansion
of the parabolic cylinder function \( D_p(z) \) as \( T\rightarrow
\infty \), which corresponds to an asymptotic expansion as the complex
order $p$ goes to \( i\infty \). Unfortunately, we do not know of any
systematic way of doing this because asymptotic expansion for $D_p$
known in the literature corresponds to either the limit
\begin{equation}
\label{eq:knownlimit1}
\sqrt{m\beta T}|\tau |\gg \lambda T/2
\end{equation}
or\begin{equation}
\label{eq:knownlimit2}
\lambda T/2\gg m\beta T\tau ^{2}\, ,
\end{equation}
both of which are not satisfactory 
for generic \( \tau  \). However,
for \( \tau \rightarrow \pm \infty  \) we 
can satisfy Eq.~(\ref{eq:knownlimit1}),
and for \( \tau \rightarrow 0 \) we can satisfy Eq.~(\ref{eq:knownlimit2}).
In reproducing the result of Audretsch and Schaeffer, we will focus
on the limit \( \tau \rightarrow \pm \infty  \) satisfying 
Eq.~(\ref{eq:knownlimit1}).

In the limit of \( \tau \rightarrow -\infty \), the parabolic cylinder
function has the asymptotic formula\begin{eqnarray} D_{p}(z) & \sim &
z^{p}e^{-z^{2}/4} \left[1-\sum _{n=1}{\cal
O}(p^{2n}/z^{2n})+...\right]\\ & = & 2^{-1/4+i\lambda T/4}e^{\pi
\lambda T/8+i\pi /8} (m\beta T)^{\frac{-1}{4}(1-i\lambda T)} \nonumber
\\ & \times & |\tau |^{-(1-i\lambda T)/2} e^{\frac{im\beta T\tau
^{2}}{2}}\left[1- \sum_{n=1}{\cal O}\left(\frac{(\lambda
T)^{2n}}{(m\beta T)^{n} \tau
^{2n}}\right)+...\right]\label{eq:Dplimit0}
\end{eqnarray}
where\begin{equation}
z=(i-1)\sqrt{m\beta T}\tau\, , 
\end{equation}
\begin{equation}
p=-(1-i\lambda T)/2\, ,
\end{equation}
giving\begin{equation}
\label{eq:parabolic}
f\sim \frac{e^{\frac{im\beta T\tau ^{2}}{2}}}{\sqrt{2m\beta 
|\tau |}}2^{i\lambda T/4}e^{i\pi /8}(m\beta T)^{i\lambda T/4}|\tau 
|^{i\lambda T/2}
\end{equation}
where one has to be careful to write\begin{equation}
(i-1)\sqrt{m\beta T}\tau =-(i-1)\sqrt{m\beta T}|\tau |
\end{equation}
 for \( \tau <0 \).\footnote{We see 
that in Eq.~(\ref{eq:Dplimit0}), we cannot take \( T\rightarrow \infty  \) 
after taking \( \tau \rightarrow -\infty  \) as the \( T \) 
expansion is out of control in that limit ({\it i.e.} \( T \) limit 
and \( \tau  \) limit do not commute unless 
the \( T \) expansion can be summed up).}

Now, we construct the template function.  Let us start with the zeroth
order expansion \begin{equation} W_{k}^{[0]}(\tau )=\sqrt{m\beta
\lambda +m^{2}\beta ^{2}\tau ^{2}}
\end{equation}
which gives\begin{equation}
h_{k}^{[0]}=\frac{[2(\sqrt{\beta m}\tau +\sqrt{\lambda +\beta m\tau
^{2}})]^{-i\lambda T/2}}{\sqrt{2\sqrt{m\beta \lambda +m^{2}\beta
^{2}\tau ^{2}}}}\exp \left[\frac{-iT}{2}\sqrt{\beta m(\lambda +\beta m\tau
^{2})}\tau \right]\, .
\label{eq:hk0p}
\end{equation}
In the limit \( \tau \rightarrow -\infty  \), we therefore have\begin{equation}
\label{eq:wkb0}
h_{k}^{[0]}\sim \frac{1}{\sqrt{2m\beta |\tau |}}
\exp \left[-i\frac{m\beta \tau |\tau |T}{2}\right]
(\sqrt{m\beta }|\tau |)^{i\lambda T/2}(\lambda )^{-i\lambda T/2}\, .
\end{equation}

We can consider higher orders: \begin{equation}
W_{k}^{[2]}=\sqrt{\beta m(\lambda +\beta m\tau ^{2})}+\frac{(3\beta 
m\tau ^{2}-2\lambda )\sqrt{\beta m(\lambda +\beta 
m\tau ^{2})}}{8T^{2}(\lambda +\beta m\tau ^{2})^{3}}\, ,
\end{equation}
\begin{eqnarray}
W^{[4]}_{k} & = & \sqrt{\beta m(\lambda +\beta m\tau ^{2})}+\frac{(3\beta m\tau ^{2}-2\lambda )\sqrt{\beta m(\lambda +\beta m\tau ^{2})}}{8T^{2}(\lambda +\beta m\tau ^{2})^{3}}\\
 &  & -\frac{(76\lambda ^{2}-732\beta \lambda m\tau ^{2}+
297\beta ^{2}m^{2}\tau ^{4})\sqrt{\beta
 m(\lambda +\beta m\tau ^{2})}}{128T^{4}(\lambda +\beta m\tau ^{2})^{6}}\, .
\end{eqnarray}
In the limit that \( \tau \rightarrow -\infty  \), these have the
expansions\begin{equation}
W_{k}^{[2]}\sim \left(\beta m|\tau |+\frac{\lambda |\tau |}{2\tau ^{2}}
-\frac{\lambda ^{2}|\tau |}{8\beta m\tau ^{4}}+...\right)+
\frac{1}{T^{2}}\left(\frac{3|\tau |}{8\beta m\tau ^{4}}+
\frac{19\lambda |\tau |}{16\beta ^{2}m^{2}\tau ^{6}}+...\right)\, ,
\end{equation}
\begin{equation}
W_{k}^{[4]}\sim W_{k}^{[2]}+\frac{1}{T^{4}}\left(\frac{-297|\tau |}{128
\beta ^{3}m^{3}\tau ^{8}}+
\frac{4731\lambda |\tau |}{256\beta ^{4}m^{4}\tau ^{10}}+...\right)
\label{eq:fourthorderspecial}
\end{equation}
where we note that each power of \( 1/T^{n} \) has an expansion in
\( \frac{\lambda }{\beta m\tau ^{2}} \) while the leading correction
coming from \( 1/T^{n} \) terms is \( |\tau |/([\beta m]^{n-1}\tau ^{2n}) \).
This means that even for a fixed finite $T$, we can write
\begin{equation}
\lim_{\tau_0 \rightarrow -\infty} W^{[A]}(\tau_0)/W^{[0]}(\tau_0) =1
\label{eq:perfectlyadiabatic}
\end{equation}
and hence we can conclude that this corresponds to an exceptional
situation in which infinite adiabaticity occurs at the time
$\tau_0=-\infty$ when the vacuum is set.  At any finite time $\tau_0$,
we would have to use the matching in expansion of $1/T$, which can
lead to nonperturbative ambiguity, but here, all the terms in $1/T$
are effectively identically 0.  For completeness, we integrate to
display the template function exponent.  We find
\begin{eqnarray}
T\int ^{\tau }dxW_{k}^{[4]}(x) & = &\left[\frac{1}{2}\beta Tm|\tau |
\tau +\frac{\lambda T|\tau |}{2\tau }\ln |\tau |+...\right]
\nonumber \\ 
& & +\frac{1}{T}\left[\frac{-|\tau |}{\tau }\frac{3}{16\beta m\tau ^{2}}+...
\right]+{\cal O}\left(T^{-3}\right)
\label{eq:wkb4phase}
\end{eqnarray}
which means that when exponentiated, both the \( 1/T \) and \( 1/\tau  \)
expansions are both well in control. 

Because of \eqr{eq:perfectlyadiabatic}, not only do
Eqs.~(\ref{eq:wkb0}) and (\ref{eq:parabolic}) match up to \( \tau \)
independent phases (which can always be chosen appropriately in the
indefinite integral Eq.~(\ref{template})), but the specification is
unique without any nonperturbative ambiguity.  Furthermore, we have
from \eqr{eq:perfectlyadiabatic} that \( h_{k}^{[0]}(\tau \sim -\infty
)=h_{k}^{[\infty ]}(\tau \sim -\infty ) \), implying that the matched
vacuum $h_k$ is an infinite adiabatic order vacuum.  We shall denote
this vacuum as
\begin{equation}
h^{\tau _{0}=-\infty }_{k}(\tau )=
\frac{1}{(2m\beta )^{1/4}}e^{-\pi
\lambda /8}D_{-(1-i\lambda )/2}[(i-1)\sqrt{m\beta }\tau ] 
\label{eq:pastvacuumwunc} 
\end{equation}
 where we have denoted the placement of the boundary condition at \(
\tau _{0}=-\infty \).

Since we can read off the template function at \( \tau =\infty \) from
Eqs.~(\ref{eq:hk0p}) and (\ref{eq:wkb0}), we can write
\begin{equation}
 h^{\tau _{1}=\infty }_{k}(\tau )=h_{k}^{\tau _{0}=-\infty }(-\tau
)^{*}
\end{equation}
specifying the vacuum at a future time.  By examining the asymptotic
expansions 9.246 of \cite{gr}, we can then write
\begin{equation}
h_{k}^{\tau_0=-\infty}(\tau >0)  =  \left[ \frac{-\sqrt{2\pi }e^{i\pi /4}
e^{-\pi \lambda /4}}{\Gamma \left[\frac{1}{2}(1-i\lambda )\right]} 
 \right]
h_{k}^{\tau _{1}=\infty }(\tau ) 
 + \left[-ie^{-\pi \lambda /2} \right]h_{k}^{\tau _{1}= \infty *}(\tau ) \, ,
\end{equation}
 giving\footnote{There is an error in the book by Birrell and Davies
\cite{birrelldavies} 
 for the first coefficient.}
a particle production coefficient of\begin{equation}
\beta_{k}=-ie^{-\pi \lambda /2}.
\label{eq:bkbounce}
\end{equation}
More explicitly, if we had not set $T=1$ through \eqr{eq:removeT} we
could write \( \beta_{k}=-ie^{-\pi T\lambda /2} \), which shows
explicitly that this coefficient is \( 0 \) to all orders in \( 1/T
\).  Therefore, this nonperturbatively suppressed particle production
could not have been reliably calculated if we had to make an expansion
in $1/T$ at any point in the calculation (we have made an expansion in
$1/T$ in this section just to show that it was not necessary as can be
seen for example in \eqr{eq:perfectlyadiabatic}).  Such an expansion
in $1/T$ would have been necessary if the time $\tau_0$ at which
vacuum is specified were finite.

\section{\label{sec:slowroll} The adiabatic vacuum in the slow roll spacetime}
For any realistic scenario of inflation, the background spacetime is
quasi-dS, unlike the exact dS that we treated explicitly before. By
quasi-dS, we mean the Hubble rate ($H=\dot{a}(\tau)/a^2(\tau)$ ) is not
exactly constant, but changes with time as $\dot{H}/a=-\epsilon H^2$,
where $\epsilon$ is one of the so-called slow-roll
parameters.\footnote{Recall we are using conformal time coordinates.}
The other useful slow-roll parameter is $\eta=\left(V^{\prime\prime}/3
H^2\right)$ where $V(\phi)$ is the inflaton potential and primes
denote derivatives with respect to the inflaton field $\phi$.  We wish
to estimate the uncertainty in the vacuum in such a spacetime, and
thereby estimate the uncertainty in the inflationary prediction of
density perturbations.

In the gauge invariant treatment of cosmological perturbations
generated during inflation, it is useful to define the gauge invariant 
Mukhanov variable $u=a\delta\phi^{\left( {\rm GI}\right)}+z\Psi$
\cite{mukhanov} where $z= \dot\phi/H$ and 
$\delta\phi^{\left( {\rm GI}\right)}$ $\Psi$ are the gauge invariant 
inflaton fluctuation and the Bardeen gravitational potential respectively 
\cite{bardeen}.  Performing for the
variable $u$ a field expansion similar to Eq.~(\ref{h}), one finds the
mode equation (after inserting the adiabatic parameter $1/T$)
\begin{equation}
\frac{1}{T^2}\ddot{\tilde{h}}_k(\tau) +\left[ k^2 -\frac{(2+p)}{T^2 
\tau^2} \right]\tilde{h}_k(\tau) = 0,
\label{modeequationSR}
\end{equation}
where $p\equiv 3 (3\epsilon-\eta)$. 
 The exact mode equation solution can be
written in terms of Hankel functions as
\begin{equation}
\tilde{h}_{k}=\sqrt{\frac{\pi }{2}}\sqrt{-\tau T}\left[ A_{k}H_{\nu
}^{(1)}(-kT\tau )e^{i\frac{\pi }{2}(\frac{1}{2}+\nu )}+B_{k}H_{\nu
}^{(2)}(-kT\tau )e^{-i\frac{\pi }{2}(\frac{1}{2}+\nu )}\right]
\label{nb}
\end{equation}
where one must be careful to note that \( -\tau >0 \) (we have also
taken out \( Tk\tau \) independent phases from \( A_{k} \) and \(
B_{k} \) for convenience and normalized conveniently). Here, the order
is defined as $\nu \equiv \sqrt{9+4p}/2$. 
\footnote{The Hankel's functions are defined as
$H^{(1)}_{\nu}=J_{\nu}+i Y_{\nu}$ and $H^{(2)}_{\nu}=H^{(1) \
*}_{\nu}$, where $J_{\nu}$ and $Y_{\nu}$ are, respectively, the Bessel
functions of the first and the second kind.}

Let us now construct the adiabatic vacuum following the recipe of
subsection \ref{adiabaticvacuumsection}.  After rescaling $d/d\tau
\rightarrow (1/T) d/d\tau$ in the mode equation, the template
frequencies can be constructed as
\begin{eqnarray}
W_k^{[0]\ 2}(\tau)&=&k^2\, , \\ 
W_k^{[2]\ 2}(\tau)&=&k^2\left[
1-\frac{(2+p)}{(k\tau T)^2} \right]\, , \\
W_k^{[4]\ 2}(\tau)&=&k^2\left[ 1-(2+p)\frac{1}{(k\tau T)^2}  +
                \left(3+\frac{3}{2}p\right)\frac{1}{(k\tau T)^4}
		+{\cal O} \left( \frac{1}{(k\tau T)^6} \right) \right]\, , \\
W_k^{[6]\ 2}(\tau)&=&k^2\left[ 1-(2+p)\frac{1}{(k\tau T)^2}  +
               \left(3+\frac{3}{2}p\right)\frac{1}{(k\tau T)^4} +
              \left(-4+\frac{7}{2}p\right)\frac{1}{(k\tau T)^6} \right. 
\nonumber\\
&+&\left. {\cal O}\left(\frac{1}{(k\tau T)^8}\right)\right] \, .
              \label{WSR}
\end{eqnarray}
However, unlike in the dS case, the series is generally non-convergent
for any finite $\tau=\tau_0$.  As we discussed before, this type of
nonconvergence is more generic than the convergent case.

To fourth order in adiabatic expansion, we find the following WKB
template function\begin{equation}
\label{eq:hkwkb}
h^{[4]}_{k}=\frac{e^{-ik\tau T}}{\sqrt{2k}}\left( 1+\frac{-i(1+p/2)}
{kT\tau }-\frac{p}{4k^{2}T^{2}\tau ^{2}}-\frac{ip}{6k^{3}T^{3}\tau ^{3}}+
\frac{5p}{24k^{4}T^{4}\tau ^{4}}\right) \, .
\end{equation}
For the
asymptotic expansion of the Hankel functions \( H_{\nu }^{(1,2)} \),
we have from Ref.~\cite{gr} 
\begin{equation}
H_{\nu }^{(1)}(z)=\sqrt{\frac{2}{\pi z}}e^{i\left(z-\frac{\pi }{2}\nu
-\frac{\pi }{4}\right)}\left[ \sum
_{k=0}^{n-1}\frac{(-1)^{k}}{(2iz)^{k}}\frac{\Gamma (\nu
+k+\frac{1}{2})}{k!\Gamma (\nu -k+\frac{1}{2})}+\theta_1
\frac{(-1)^n}{\left(2iz\right)^n}\frac{\Gamma\left(
\nu+n+\frac{1}{2}
\right)}{\Gamma\left(
\nu-n+\frac{1}{2}
\right)} \right] \, ,
\label{eq:hankel1}
\end{equation}
\begin{equation}
H_{\nu }^{(2)}(z)=\sqrt{\frac{2}{\pi z}}e^{-i\left(z-\frac{\pi }{2}\nu
-\frac{\pi }{4}\right)}\left[ \sum
_{k=0}^{n-1}\frac{1}{(2iz)^{k}}\frac{\Gamma (\nu
+k+\frac{1}{2})}{k!\Gamma (\nu -k+\frac{1}{2})}
 +\theta_2
\frac{1}{\left(2iz\right)^n}\frac{\Gamma\left(
\nu+n+\frac{1}{2}
\right)}{\Gamma\left(
\nu-n+\frac{1}{2}
\right)}\right] \, ,
\label{eq:hankel2}
\end{equation}
where \( \textrm{Re}(\nu )>-1/2 \), \( |\textrm{arg}(z)|<\pi  \)
and $\theta_{1,2}$ are coefficients smaller than unity in front of the
remainders.
A useful formula is\begin{equation}
\frac{\Gamma (\nu +k+\frac{1}{2})}{\Gamma (\nu -k+\frac{1}{2})}
=\frac{(4\nu ^{2}-1^{2})(4\nu ^{2}-3^{2})...(4\nu ^{2}-(2k-1)^{2})}{2^{2k}}\, .
\end{equation}
Expanding to first order in \( p \) and fourth order in \( T \),
one can then compute \begin{eqnarray}
\tilde{h}_{k}(\tau ) & \approx & \frac{1}{\sqrt{2k}}\left[ 
(A_{k}f+B_{k}f^{*})+\frac{i}{k\tau T}\{-A_{k}\left(1+\frac{p}{2}\right)
f+B_{k}\left(1+\frac{p}{2}\right)f^{*}\}\right. \nonumber \\
 & -& \frac{p}{4(k\tau T)^{2}}\{A_{k}f+B_{k}f^{*}\}+\frac{ip}{6(k\tau T)^{3}}\{-A_{k}f+B_{k}f^{*}\}\nonumber \\
 &+& \left. \frac{5p}{24(k\tau T)^{4}}\{A_{k}f+B_{k}f^{*}\}\right]\, ,
 \label{eq:hkexact} 
\end{eqnarray}
where $f\equiv e^{-ik\tau T}$.
Matching \( \tilde{h}_{k} \) and \( h_{k}^{[4]} \) just by inspection (comparing
Eqs.~(\ref{eq:hkwkb}) and (\ref{eq:hkexact}) ), one sees that $A_k=1$
and $B_{k}=0$ as expected.

Just to check further, suppose we took the 4th order template \(
h_{k}^{[4]} \) and expanded to 6th order in \( T^{-1} \) and matched
it to \( \tilde{h}_{k} \) to 6th order (instead of just to 4th order). To
accomplish this, note that since \( W^{[4]} \) only needs to be
expanded to \( T^{-6} \) accuracy even though it is being multiplied
by \( T \) since the next order term in \( W^{[4]} \) expansion is \(
T^{-8} \) , resulting in a correction of \( T^{-7} \) which we are
ignoring anyway. We find
\begin{equation} h^{[4]}_{k(6{\rm th})}=h_{k}^{[4]}+\frac{e^{-ik\tau T
}}{\sqrt{2k}}\left[\frac{-3i}{8k^{5}\tau
^{5}T^{5}}(4+3p)+\frac{9+5p}{4k^{6}\tau ^{6}T^{6}}\right]\ .
\end{equation}
This results in\begin{equation}
A_{k}=1 + \frac{3(2+p)i}{4k^{5}T^{5}\tau _{0}^{5}}+{\cal O}(T^{-6})\, ,
\end{equation}
\begin{equation}
B_{k}=\frac{-15(2+p)}{8k^{6}T^{6}\tau _{0}^{6}}e^{-2ikT\tau _{0}}\, ,
\end{equation}
where \( \tau _{0} \) is the time at which the boundary conditions
are placed.  We see explicitly the order of the residual corrections
to the 4th order asymptotic expansion, and it is as expected.
One can easily check similarly that the 6th adiabatic order WKB
template function also results in $A_{k}=1$ and $ B_{k}=0$ to 6th
order in $1/T$.  More explicitly, we have the template function
\begin{equation}
h_{k}^{[6]}=h_{k}^{[4]}+\frac{e^{-ik\tau
}}{\sqrt{2k}}\left[\frac{3ip}{8k^{5}\tau ^{5}T^{5}}-\frac{7p}{8k^{6}\tau
^{6}T^{6}}\right]
\end{equation}
that fixes the vacuum.  Indeed, the fact $A_k=1$ and $B_k=0$ is no
surprise since the formulae Eqs.~(\ref{eq:hankel1}) and
(\ref{eq:hankel2}) and the template functions are the same asymptotic
expansions to any given order. 

The matching at time $\tau_0$ and at any given adiabatic order $A$ of
the template function $h_{k}^{[A]}$ with the mode function (\ref{nb})
suffers of an ambiguity introduced by the nonvanishing remainders
(proportional to $\theta_i$) displayed in expansions
(\ref{eq:hankel1}) and (\ref{eq:hankel2}).  As discussed earlier, the
order at which the remainder stops converging (with $T=1$) can be
attributed to be the extent to which the vacuum is nonadiabatic.
Hence, instead of a guessed nonperturbative uncertainty of
\eqr{eq:bkexp}, we can set $T=1$ and compute the uncertainty that is
due to the intrinsic nonadiabaticity of the background spacetime.
(This is something we discussed at the end of
Sec.~\ref{stopsconverging}.)

More explicitly, we can look for the adiabatic order $n_*$ at which
the (absolute values of the) remainders are minimized, and then say
that we cannot define the vacuum more certainly than this order due to
the nonadiabaticity. Expanding the remainders to first-order in the
slow-roll parameters, we can express the absolute value of the
remainder as
\begin{equation}
R(n)\equiv \left|\frac{(-1)^n \Gamma (\nu+n+1/2)}{(2 i k|\tau|)^n
\Gamma(\nu-n+1/2)} \right| \approx \frac{p \Gamma(2+n)
\Gamma(n-1)}{3\cdot 2^n (k |\tau_0|)^n}  + {\cal O}(p^2) \, .
\end{equation}
We see clearly that the remainder vanishes in the dS limit of $p=0$,
which corresponds to $\epsilon=\eta=0$. This is
another way of understanding why in the pure de Sitter case we have
been able to construct a convergent $W^{[\infty]}$.  At a formal
level, it merely means that the asymptotic expansion point $1/T
\rightarrow 0$ is an analytic point, allowing a convergent asymptotic
expansion.  

We can find the minimum of $R$ at the location by taking a discrete
derivative $R(n_*+1)-R(n_*)=0$.  Although $n_*$ will not lie at an integer
value in general, the actual solution will be at the nearest
integer.\footnote{When solving, however, one will find it convenient
to assume self-consistently that $n_*$ will in the end be at an integer
value.}  We thus find
\begin{equation}
n_* = \mbox{integer}\left(\frac{-1}{2} +
\frac{1}{2}\sqrt{1+8(1+k|\tau_0|)} \right)\approx \mbox{integer}
\left(\sqrt{2 k  |\tau_0|}\right)
\end{equation}
to be the order at which the uncertainty is minimized where the
$\mbox{integer}$ function finds the nearest integer value of the
argument.  Note that the value is independent of $p$ to leading order
because in the limit that $p$ vanishes, the corrections identically
vanish, giving an ``arbitrary'' $n_*$.  In terms of e-folds, we can
write
\begin{equation}
n_*\simeq \sqrt{2 } \exp((N_0 - N_k)/2)
\end{equation}
where $N_k$ denotes the number of $e$-foldings from the time when a
given wavelength $\lambda=a/k$ leaves the horizon ($\lambda=1/H$) till
the end of inflation and $N_0$ denotes the number of $e$-foldings from
the time the vacuum was set to the end of inflation. As we already
mentioned, length scales of interest for the the CMB anisotropies give
$N_k$ of order of 60. 

The corresponding ambiguity in the value of the parameter $B_k$ can be
calculated from $R(n)$ as follows.  We can consider the template
function to be
\begin{equation}
h_k^{[n_*-1]} \equiv \sqrt{\frac{\pi |\tau| }{2}} H_\nu^{(1)} (-k\tau) e^{i
(\frac{1}{2} +\nu )\frac{\pi}{2}} + r_{n_*} \, ,
\end{equation}
where $r_n$ is related to the remainder as 
\begin{eqnarray}
r_{n} & = & h_k^{[0]} \tilde{R}_1(n)\, , \\
h_k^{[0]} & = & \frac{e^{-i k \tau}}{\sqrt{2k}}\, , \\
\tilde{R}_1(n) & = &\frac{\sqrt{2} (-1)^n \Gamma(\nu + n+ \frac{1}{2})
e^{ -i (\frac{\pi}{2} \nu +  
\frac{\pi}{4})}}{(2 i |\tau| k)^n \Gamma(\nu - n +\frac{1}{2}) } \, ,
\end{eqnarray}
where $|\tilde{R}_1(n)| =\sqrt{2} R(n)$.
Now, solving for the coefficients $A_k$ and $B_k$ using the system 
\begin{eqnarray}
h_k^{[n_*-1]}(\tau_0) & = & \tilde{h}_k(\tau_0)|_{T=1} \, , \\
h_k^{[n_*-1]'}(\tau_0) & = & \tilde{h}_k'(\tau_0)|_{T=1}  
\end{eqnarray}
gives
\begin{equation}
B_k \approx -i h^{[0]2} \frac{d\tilde{R}_{1}(n_*)}{d\tau}\, .
\end{equation}
The ambiguity in the power spectrum ${\cal P}_{\cal R}$ of the
comoving curvature perturbation ${\cal R}=u/z$ is given by
\begin{equation}
\left|\frac{\delta{\cal P}_{\cal R}(k)}{{\cal P}_{\cal
R}(k)}\right|\simeq 2\, \left| {\rm Re} \, B_k \right| \simeq {\cal O} \left[
p \,  \exp( -2\sqrt{2} e^{ (N_0 - N_k)/2 } - (N_0 - N_k)/2) \right] \, ,
\label{eq:betterspecunc}
\end{equation}
where we are using the notation and approximation explained below
\eqr{eq:bunchdaviesspecunc}.  Note that this is generically slightly
larger than \eqr{eq:bkexpinton} and that it artificially vanishes in
the limit that $p\rightarrow 0$.  In the case that $p\rightarrow 0$,
the best estimate that we can give for the uncertainty is
\eqr{eq:bkexpinton}, which is a merely a guessed function that drops
off nonperturbatively fast. The theoretical ambiguity on the power
spectrum of the comoving curvature perturbation is sizeable if the
total duration of the de Sitter stage corresponds to a number of
$e$-foldings not far from 60.  Of course, in this regime, it may not
be a good approximation to treat the spacetime to be void of
fluctuations.

\section{Conclusions}
Within the context of Bunch-Davies vacuum formalism and the adiabatic
vacuum formalism we have answered the following question: ``What is
the minimal theoretical uncertainty in the inflationary curvature
perturbation calculation if we assume that the curvature perturbation
state at sometime $\tau_0$ near the beginning of inflation was a slow
roll vacuum?''  Even without any trans-Planckian effects, effective
field theory cutoff related effects, or nongravitational field
interaction effects, there is a minimal uncertainty in the curvature
perturbation predictions of inflation coming from the inability to
uniquely specify a vacuum (due to gravitational interactions).  Within
the adiabatic vacuum formalism, the power spectrum uncertainty is
given by \eqr{eq:betterspecunc}, and in more general situations
(applicable to outside of the inflationary regime) by \eqr{eq:bkexp}.
The Bunch-Davies formalism gives a larger uncertainty of
\eqr{eq:bunchdaviesspecunc}.

The minimum uncertainty presented here applies to all of the efforts
\cite{transpl,danielsson,Burgess:2002ub,Maldacena:2002vr} to obtain
measurable small effects on the CMB.  In practice of comparing with
data, there will certainly be other theoretical uncertainties, not
only from other interaction effects of the inflaton, but reheating
historical uncertainties \cite{Feng:2003nt} as well as astrophysical
uncertainties which will most likely overwhelm the minimal uncertainty
presented here, unless the number of $e$-foldings is very close to the
minimum required for inflation (e.g. around 60).  Even in that case,
however, most likely, it will be difficult to assume that the
curvature perturbation quantum state is that of a vacuum due to other
energy density fluctuations present which depend on the history
leading to the initiation of inflation.  Nonetheless, it is important
and reassuring to know that the inflationary vacuum in the
conservative sense has very little ambiguity contrary to the
impressions given particularly by Ref.~\cite{danielsson}.  Indeed, the
main qualitative conclusion one can draw from our work is that even
with a finite period of inflation, we can in most cases neglect the
uncertainty associated with the vacuum when defined according to the
most reasonable particle based definitions.

Finally, we would like to comment that the general idea that the
vacuum is uncertain with a finite period of inflation is not entirely
new (see for example \cite{fulling,Dolgov:1994ra}).  Here, we merely
tried to quantify this in the context of slow roll inflationary models
and show explicitly that the best estimates for the uncertainty lead
us not to worry about this effect unless the total duration of
inflation is close to 60 e-folds or so.

\vspace{24pt}
\centerline{\bf ACKNOWLEDGMENTS}
\vspace{24pt}
\noindent
We would like to thank J.~Cline, M.~Einhorn, L.~Everett, E.~Kolb,
F.~Larsen, R.~Leigh, L.~Parker, R.~Rattazzi, J.~Van der Schaar,
L.~Senatore and I.~Tkachev for conversations regarding this topic.  We
would like to thank in particular M.~Einhorn and L.~Parker for their
comments on an early manuscript of this work.

\frenchspacing
\def\prpts#1#2#3{Phys. Reports {\bf #1}, #2 (#3)}
\def\prl#1#2#3{Phys. Rev. Lett. {\bf #1}, #2 (#3)}
\def\prd#1#2#3{Phys. Rev. D {\bf #1}, #2 (#3)}
\def\plb#1#2#3{Phys. Lett. {\bf #1B}, #2 (#3)}
\def\npb#1#2#3{Nucl. Phys. {\bf B#1}, #2 (#3)}
\def\apj#1#2#3{Astrophys. J. {\bf #1}, #2 (#3)}
\def\apjl#1#2#3{Astrophys. J. Lett. {\bf #1}, #2 (#3)}
\begin{picture}(400,50)(0,0)
\put (50,0){\line(350,0){300}}
\end{picture}

\end{document}